\documentclass[utf8]{FrontiersinHarvard} % for articles in journals using the Harvard Referencing Style (Author-Date), for Frontiers Reference Styles by Journal: https://zendesk.frontiersin.org/hc/en-us/articles/360017860337-Frontiers-Reference-Styles-by-Journal
\usepackage{url,hyperref,lineno,microtype,subcaption}
\usepackage[onehalfspacing]{setspace}

\newcommand{\pt}{$p_\mathrm{T}\ $}
\def\keyFont{\fontsize{8}{11}\helveticabold }
\def\firstAuthorLast{Pelayo Leguina {et~al.}} %use et al only if is more than 1 author
\def\Authors{Pelayo Leguina$^{1,*}$ and Santiago Folgueras\,$^{1}$}
\def\Address{$^{1}$Department of Physics and Institute of Space Sciences and Technologies of Asturias (ICTEA), University of Oviedo, Spain}
\def\corrAuthor{Pelayo Leguina L\'opez}
\def\corrEmail{leguinapelayo@uniovi.es}

\begin{document}
\onecolumn
\firstpage{1}

\title[Hardware Acceleration in High-Energy Physics through HLS]{Harnessing Hardware Acceleration in High-Energy Physics through High-Level Synthesis Techniques} 

\author[\firstAuthorLast ]{\Authors} %This field will be automatically populated
\address{} %This field will be automatically populated
\correspondance{} %This field will be automatically populated

\extraAuth{}% If there are more than 1 corresponding author, comment this line and uncomment the next one.
%\extraAuth{corresponding Author2 \\ Laboratory X2, Institute X2, Department X2, Organization X2, Street X2, City X2 , State XX2 (only USA, Canada and Australia), Zip Code2, X2 Country X2, email2@uni2.edu}

\maketitle

\begin{abstract}

%%% Leave the Abstract empty if your article does not require one, please see the Summary Table for full details.
\section{}

At the Large Hadron Collider, the vast amount of data from experiments demands not only sophisticated algorithms but also substantial computational power for efficient processing. This paper introduces hardware acceleration as an essential advancement for high-energy physics data analysis, focusing specifically on the application of High-Level Synthesis (HLS) to bridge the gap between complex software algorithms and their hardware implementation. We will explore how HLS facilitates the direct implementation of software algorithms into hardware platforms such as FPGAs, enhancing processing speeds and enabling real-time data analysis. This will be highlighted through the case study of a track-finding algorithm for muon reconstruction with the CMS experiment, demonstrating HLS’s role in translating computational tasks into high-speed, low-latency hardware solutions for particle detection and reconstruction. Key techniques in HLS, including parallel processing, pipelining, and memory optimization, will be discussed, illustrating how they contribute to the efficient acceleration of algorithms in high-energy physics. We will also cover design methodologies and iterative processes in HLS to optimize performance and resource utilization, alongside a brief mention of additional techniques like algorithm approximation and hardware/software co-design. In short, this paper will underscore the potential of hardware acceleration in high-energy physics research, emphasizing HLS as a powerful tool for physicists to enhance computational efficiency and foster groundbreaking discoveries.

\tiny
 \keyFont{ \section{Keywords:} High-Level Synthesis (HLS), Hardware Acceleration, Field-Programmable Gate Arrays (FPGAs), Muon Reconstruction, CMS Experiment, Track-Finding Algorithm, High-Energy Physics Data Analysis, Parallel Processing and Pipelining} %All article types: you may provide up to 8 keywords; at least 5 are mandatory.
\end{abstract}

\section{Introduction}

The Large Hadron Collider (LHC) at CERN represents the pinnacle of high-energy physics (HEP) research, enabling scientists to probe the fundamental constituents of matter and the forces governing their interactions. Experiments like the Compact Muon Solenoid (CMS) generate an unprecedented volume of data, with collision events occurring at rates of several billions per second~\citep{Evans_2008}. This deluge of data necessitates not only sophisticated algorithms for accurate particle detection and reconstruction but also demands substantial computational resources to process the information in real time.

Traditional software-based data processing approaches, while flexible, often struggle to meet the low-latency and high-throughput requirements of modern HEP experiments. The latency constraints are particularly stringent in trigger systems, where rapid decision-making is crucial to determine which events are of interest and should be recorded for further analysis~\citep{CERN-LHCC-2020-004}. Hardware acceleration emerges as a vital solution to these challenges, offering significant improvements in processing speeds and enabling real-time data analysis. 

In the context of the CMS experiment, fast and efficient particle reconstruction is essential for having a reliable trigger system and for the success of the physics program~\citep{Radburn2022}. Implementing track-finding algorithms directly into hardware accelerators can drastically improve processing speeds and reduce latency. Previous efforts have demonstrated the feasibility of using FPGAs for real-time tracking in HEP experiments~\citep{ATLAS:2021tfo,Tomalin:2017hts}. However, the manual translation of algorithms into hardware description languages is time-consuming and error-prone.

High-Level synthesis (HLS) has gained traction as a powerful tool that bridges the gap between complex software algorithms and their hardware implementations. HLS allows for the description of hardware functionality using high-level programming languages like C or C++, which are then synthesized into hardware description languages suitable for implementation on Field-Programmable Gate Arrays (FPGAs) or Application-Specific Integrated Circuits (ASICs)~\citep{Nane2020}. This approach significantly reduces development time and makes hardware acceleration more accessible to software engineers and physicists who may not be experts in hardware design.

\begin{figure}[h!]
\begin{center}
\includegraphics[width=\linewidth]{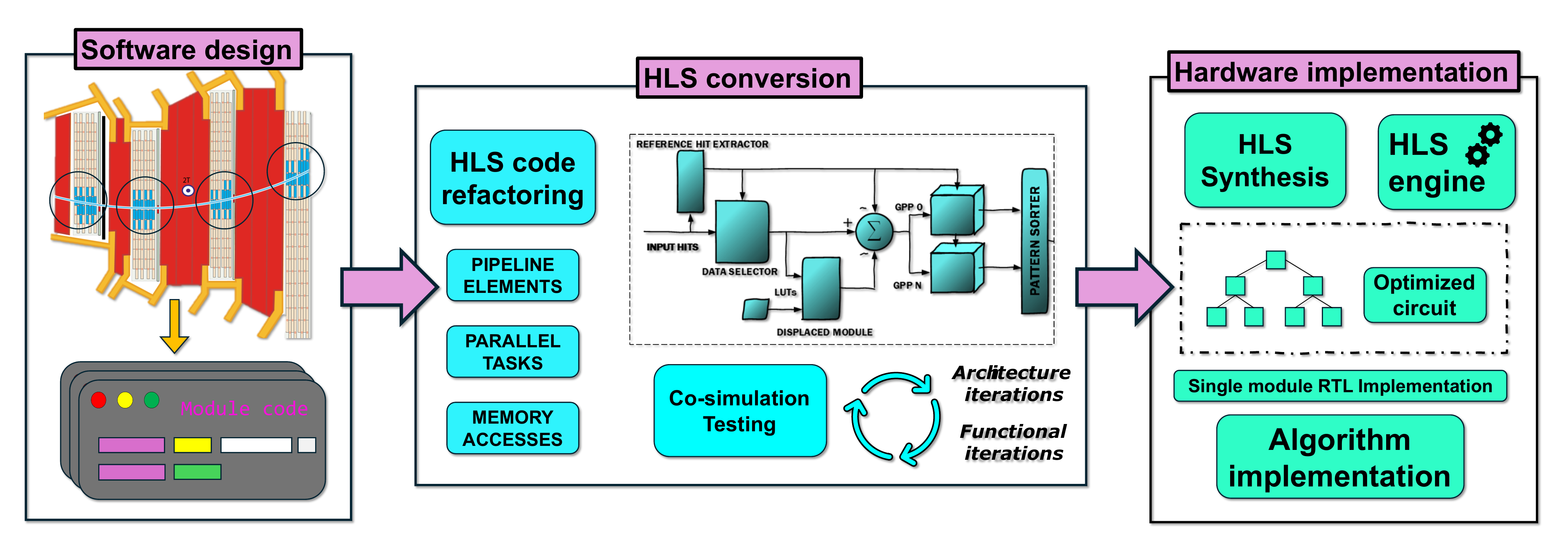}
\end{center}
\caption{The design process of high energy physics algorithms with HLS: software design, HLS refactoring, and hardware implementation.}\label{fig:init}
\end{figure}

With HLS, we can directly implement sophisticated track-finding algorithms into hardware, leveraging techniques such as parallel processing and pipelining to optimize performance (see Figure \ref{fig:init}). Parallel processing allows multiple computations to occur simultaneously, significantly increasing throughput~\citep{Nane2020}. Several studies at CERN have employed HLS to optimize pipelining and memory usage. Pipelining enables overlapping of operations, reducing the overall processing time per event, while memory optimization techniques enhance performance by minimizing access times and efficiently utilizing on-chip resources~\citep{HUSEJKO_2015,Ghanathe:2017dpm}.

This paper presents a comprehensive study on the application of HLS for hardware acceleration in HEP data analysis. We focus on the implementation of a track-finding algorithm for muon reconstruction within the CMS experiment as a case study. Our work demonstrates how HLS facilitates the translation of complex computational tasks into high-speed, low-latency hardware solutions.

We also discuss design methodologies and iterative processes in HLS to optimize performance and resource utilization. The importance of algorithm approximation is highlighted, where trade-offs between precision and computational efficiency are considered~\citep{Han2016}. In addition, we show an automated local pipeline for HLS module building, simulation, and hardware implementation, demonstrating how automation can streamline the development process and enhance efficiency.

This paper is organized as follows: Section~\ref{sec:experimental} provides an introduction to the CMS Level-1 trigger system and the software tools utilized in this work; Section~\ref{sec:AccelerationWithHLS} describes the required steps to use HLS for algorithm acceleration; Section~\ref{sec:automation} depicts the bases for an automated workflow for prototyping, validation and integration; Section~\ref{sec:results} describes the obtained results. Finally, Section~\ref{sec:discussion} summarizes our findings. 

\section{Experimental setup} \label{sec:experimental}

\subsection{The Overlap Muon Track Finder of the CMS Level-1 trigger system}

%\subsubsection{Field-programmable gate arrays}

%For the implementation of the muon track-finding algorithm using high-level synthesis, we utilized Xilinx Virtex UltraScale+ FPGAs, specifically the VU13P model. These FPGAs offer a high number of logic cells, digital signal processing (DSP) slices, and block RAMs, making them suitable for the complex, high-throughput applications required in HEP experiments. The VU13P features over 3 million logic cells, 12,288 DSP slices, and 360 Mb of UltraRAM, providing ample resources for implementing deep pipelining and parallel processing techniques inherent in HLS-optimized designs.

%The choice of FPGA allows for real-time data processing with low latency, which is critical in trigger systems like that of the CMS experiment. The reconfigurable nature of FPGAs also offers flexibility in updating and optimizing the algorithm as needed without the need for hardware changes.

%\begin{figure}[h!]
%\begin{center}
%\includegraphics[width=0.5\linewidth]{figures/x2o.png}
%\end{center}
%\caption{The custom X2O ATCA board equipped with two %Virtex UltraScale+ VU13P FPGAs.}\label{fig:x2o}
%\end{figure}

%\subsubsection{CMS Level-1 trigger system}
The CMS Level-1 trigger system will undergo a significant upgrade to accommodate the increased luminosity and data rates expected from the High-Luminosity LHC (HL-LHC)~\citep{CERN-LHCC-2020-004}. The upgraded trigger system, also referred as Phase-2 Level-1 trigger, is designed to handle up to 750 kHz of event rate with a latency of approximately 12.5 $\mu$s. Our hardware implementation integrates with this upgraded trigger system, requiring compatibility with its stringent latency constraints and data formats. A detailed description of the Phase-2 Level-1 trigger system can be found in \citep{CERN-LHCC-2020-004}.

The Phase-2 Level-1 trigger system comprises advanced electronics, including state-of-the-art FPGAs  and high-speed optical links. The system employs the Advanced Telecommunications Computing Architecture (ATCA) standards, providing high-density, high-throughput, and low-latency data processing capabilities necessary for real-time event selection in the HL-LHC environment. As a baseline, the Xilinx VU13P FPGA is used, these FPGA offers a high number of logic cells, digital signal processing (DSP) slices, and block RAMs, making them suitable for the complex, high-throughput applications required in HEP experiments. The VU13P features over 3 million logic cells, 12288 DSP slices, and 360 Mb of UltraRAM, providing ample resources for implementing deep pipelining and parallel processing techniques inherent in HLS-optimized designs.

%\begin{figure}[h!]
%\begin{center}
%\includegraphics[width=0.7\linewidth]{figures/2_figure.png}
%\end{center}
%\caption{Functional diagram of the CMS level-1 phase-2 upgraded trigger system~\citep{CERN-LHCC-2020-004}.}\label{fig:trigger}
%\end{figure}

As a case study, the track finding algorithm for the overlap muon track finder (OMTF) of the CMS experiment is used. The OMTF is designed to reconstruct muon trajectories in the barrel-endcap transition region of the detector~\citep{Zabolotny_2016}. This region is extremely complex due to the inhomogeneous magnetic field in such region and the  different geometric orientations of the muon detectors: Drift Tubes (DT), Cathode Strip Chambers (CSC), and Resistive Plate Chambers (RPC). The OMTF algorithm tackles this challenge by evaluating how well the detected hits, or stubs, correspond to expected patterns of muon tracks with specific transverse momenta (\pt). The algorithm matches the incoming detector stubs to a set of predefined reference patterns. Each pattern represents the typical configuration for a muon with a certain \pt and charge traversing the detector layers. By calculating a similarity score between the observed stubs and these reference patterns, the algorithm identifies the most probable track candidate, assigning appropriate kinematic parameters to the  muon.

Earlier implementations of the OMTF algorithm were realized in VHDL~\citep{Zabolotny_2016} and later using high-level synthesis techniques~\citep{zatbolony_hls}. Recent developments incorporated the reconstruction of displaced muons~\citep{Leguina_2023}. Our implementation incorporates optimizations to reduce latency and resource usage. By utilizing HLS, we achieve a modular and maintainable design that can be easily adapted for future upgrades and more complex detector conditions. 

In the current foreseen scenario for the HL-LHC upgrade, the OMTF system consists of 6 ATCA boards, 3 for each side of the detector. Each board should process the information from a 120º $\phi$-sector of the detector and will receive up to 1033 Gbps of information from the muon detectors. For each clock cycle, the algorithm can deliver up to 9 muons, with a maximum bandwidth of 450 Gbps (divided into 18 optical links) that will be sent to later stages of the trigger. The whole OMTF reconstruction has to be accomodate in a 2 microsecond latency budget. 

Our implementation ensures compliance with the system's timing budget and interfaces~\citep{CERN-LHCC-2020-004}. The hardware modules are designed to meet the physical, power, and thermal constraints of the CMS trigger crates, ensuring reliable operation under the demanding conditions of the HL-LHC.

%\begin{figure}[h!]
%\begin{center}
%\includegraphics[width=\linewidth]{figures/muonsys_phase2.pdf}
%\end{center}
%\caption{Longitudinal schematic view of a quarter section of the CMS detector, highlighting the muon detectors: Drift Tubes (DT, labeled as MB), Cathode Strip Chambers (CSC, ME), Resistive Plate Chambers (RPC, denoted as RB and RE), and Gas Electron Multipliers (GEM, GE). The area covered by the OMTF algorithm is also indicated.}\label{fig:omtf_section}
%\end{figure}

The dataset used for this study was generated using a trigger emulator of the aforementioned algorithm written in C++ that generates a test vector formatted in XML files consisting of a small sample of 1000 events, coming from a muon gun sample with pairs of muons with a flat \pt distribution between 1 and 100 GeV. These test vectors represent the digitized detector signals and are structured to match the input specifications of the hardware implementation. Using XML allows for a flexible and human-readable format that facilitates debugging and verification processes.

\subsection{Software Tools}
We employ AMD Vitis HLS version 2023.2 for converting the high-level algorithm descriptions into hardware description language (HDL) code suitable for FPGA implementation. Vitis HLS allows for the synthesis of C, C++, and SystemC code into Verilog or VHDL, facilitating rapid prototyping and optimization. The tool supports various optimization directives, such as loop unrolling, pipelining, and dataflow, which are essential for enhancing performance and resource utilization in FPGA designs.

The use of HLS accelerates the development cycle by enabling software engineers and physicists to design hardware accelerators using familiar programming languages. It also allows for quick iterations and testing of different optimization strategies to meet the stringent performance requirements of the CMS Phase-2 trigger system.

In addition, scripting languages such as bash and tcl were utilized for automating the build process, simulation, and generating an automated local pipeline for HLS module building, simulation, and hardware implementation. Python scripts were also employed for data analysis, visualization, and further automation tasks. The integration of these scripting languages streamlines the development workflow and facilitates collaboration among team members.

VHDL was used for lower-level hardware description and for integrating the HLS-generated modules into the existing hardware infrastructure. The use of VHDL allows for precise control over hardware resources and timing, which is critical in matching the performance requirements of the CMS Level-1 trigger system.

\section{Algorithm acceleration using high-level synthesis in High-Frequency Applications} \label{sec:AccelerationWithHLS}

The adaptation of an algorithm for hardware acceleration using high-level synthesis involves a thorough understanding of its main components and how data flows through them. The algorithm must be adapted to operate efficiently on FPGA hardware, which requires consideration of data streaming, parallel processing, and pipelining.

\subsection{Algorithm implementation}

\begin{figure}[h!]
\begin{center}
\includegraphics[width=\linewidth]{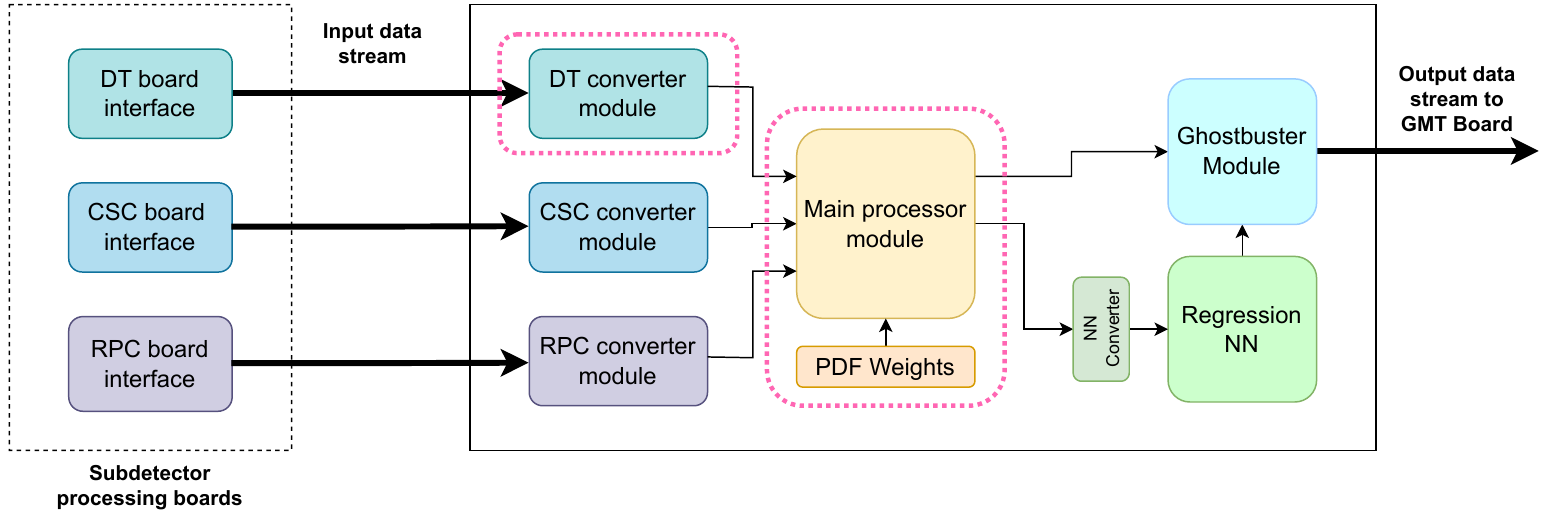}
\end{center}
\caption{Schematic block diagram of the components of the OMTF algorithm. Data from each subdetector interface flows through various modules to the Global Muon Trigger (GMT) interface. This work considers the DT converter and the main processor modules as an example to illustrate the applied high-level synthesis techniques.}\label{fig:omtf_scheme}
\end{figure}

Figure \ref{fig:omtf_scheme} shows the main modules in the HLS implementation of the OMTF algorithm, which receives input data from the multiple muon subdetectors: DT, CSC and RPC. Each produces data in its own specific format and frame structure. These data are transmitted in a streamed fashion from various parts of the Level-1 trigger system to the conversion modules of the algorithm. 

To handle this heterogeneous and continuous stream of data, we design HLS modules capable of processing streamed inputs using the \texttt{hls::stream} library. This library provides FIFO-based data streams that facilitate the handling of sequential data in a pipelined manner. The input converting modules continuously read incoming data streams, process them in real-time, and ensure that no data is lost due to buffer overflows or processing delays.

The design process of the OMTF system consists of several key steps, with a focus on efficient data handling and processing in an FPGA environment:

\begin{itemize}
    \item \textbf{Input converting modules:} These modules handle data from each subdetector using \texttt{hls::stream} objects to efficiently manage streamed data. Each module parses incoming data frames, extracts relevant hit information (primitives), and converts the data into a unified format that is compatible with the main processing module. Additionally, they merge information from multiple detector layers (DTs, CSCs, and RPCs) and ensure that the data format matches the input structure required by the main module. To minimize latency and maximize throughput, directives such as \texttt{\#pragma HLS PIPELINE} and \texttt{\#pragma HLS DATAFLOW} are used.

    \item \textbf{Main processing module:} This module receives the unified data frame through custom input ports and implements the core OMTF algorithm to infer \pt based on the spatial positions of detected primitives. In addition, the module processes multiple ``golden patterns"~\citep{Zabolotny_2016} in parallel, which represent the expected hit configurations for muons with specific \pt and charge. The design leverages \texttt{\#pragma HLS UNROLL} to unroll loops and create multiple instances of the processing logic for parallel computation. It also employs \texttt{\#pragma HLS ARRAY\_RESHAPE} to ensure that array elements storing pattern (PDF) weights and parameters are available simultaneously, eliminating data dependencies and improving efficiency. A more detailed description of this module can be found in~\citep{Zabolotny_2016}. 
\end{itemize}

This structured approach ensures that the OMTF system efficiently processes data while optimizing FPGA resources and maintaining the necessary throughput for real-time applications that was described in Section~\ref{sec:experimental}.   

Throughout the design process, we continuously verify the functionality and performance of each module using C++ test benches and HLS simulations. This iterative approach allowed us to identify bottlenecks and optimize the design before synthesizing it onto the FPGA hardware. In this work the DT converter and the processor modules are taken as an example to illustrate the applied high-level synthesis techniques.

%By translating the software algorithm into an HLS-compatible format and carefully optimizing for hardware constraints, we prepared the algorithm for efficient implementation on the FPGA platform. This preparation was crucial for meeting the stringent latency and throughput requirements of the CMS Level-1 trigger system during the HL-LHC operations.

\subsection{Optimization techniques}

Optimization was a critical aspect of our implementation to meet the stringent performance requirements of the CMS Level-1 trigger system. Several techniques were applied, focusing on parallel processing, pipelining, and memory optimization, each tailored to the specific needs of different modules within the algorithm.

\subsubsection{Parallel processing and memory optimization}

Parallel processing was primarily employed in the main processing module, where the evaluation of  patterns is inherently parallelizable. Each  pattern represents a potential muon trajectory with a specific \pt and charge. To process all patterns simultaneously, we unroll loops that iterate over the patterns using the \texttt{\#pragma HLS UNROLL} directive. By fully unrolling these loops, separate processing elements for each pattern were instantiated, allowing the module to compare incoming primitives against all patterns in a single clock cycle, significantly reducing latency as shown in Section~\ref{sec:results}.

Additionally, arrays containing pattern weights and parameters were reshaped using the \texttt{\#pragma HLS ARRAY\_RESHAPE} directive with the ``complete'' option, ensuring that all array elements were accessible in parallel without causing memory access conflicts. Constant data, such as pattern weights, were stored in read-only memory blocks, mapped to on-chip block RAMs or UltraRAMs using the \texttt{\#pragma HLS RESOURCE} directive. Array partitioning through the \texttt{\#pragma HLS ARRAY\_PARTITION} directive enabled multiple simultaneous read accesses, further optimizing memory access.

In the input converting modules, parallel processing was also applied by designing separate modules for each muon subdetector. Each module operates independently, handling its specific data stream and merging results into a unified data frame for the main module, ensuring seamless integration.

\begin{figure}[h!]
\begin{center}
\includegraphics[width=0.8\linewidth]{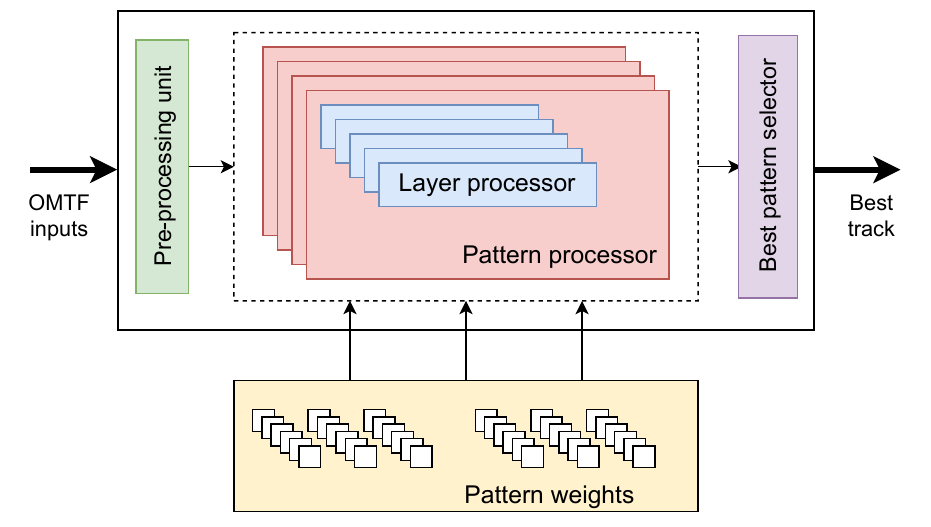}
\end{center}
\caption{Scheme of the units forming the main processor. The individualized pattern weights make full parallelization of the pattern processors available, reducing the overall latency. Each pattern processor, then processes information from each detector layer available (layer processor).}\label{fig:omtf_processor}
\end{figure}

As shown in Figure~\ref{fig:omtf_processor}, the individualized pattern weights allow full parallelization of the pattern processors, significantly reducing the overall latency of the system.

\subsubsection{Pipelining}

Pipelining was a crucial optimization technique applied to the input converting modules. Taking the DT converting module as an example, the DT module processes data frames representing muon trigger primitives, or stubs. Each event consists of 9 data frames, each 64 bits wide. Traditionally, these frames would be deserialized, combined, and processed together, introducing latency and increasing resource usage. To counter this, we use the \texttt{hls::stream} library to process primitives as they arrived, enabling a pipelined approach.

The pipeline in the DT module consists of several stages:
\begin{enumerate}
    \item \textbf{Streaming input read}: Primitives are read from the \texttt{hls::stream} input one at a time.
    \item \textbf{Quality and position filtering}: Apply detector-specific quality criteria to filter out spurious hits and ensure the hit positions are within range for muon candidates.
    \item \textbf{Coordinate conversion}: Use DSP units for local coordinate computations, applying fixed-point arithmetic to maintain precision while optimizing resources.
    \item \textbf{Data packing}: Once all 9 primitives are processed, the converted data is packed into the expected input format for the main processing module.
\end{enumerate}

To further enhance the efficiency of the DT module pipeline, the \texttt{\#pragma HLS DATAFLOW} directive was used. This directive allowed different stages of the pipeline (e.g., reading, filtering, conversion, and packing) to operate concurrently, rather than sequentially. By enabling parallel execution of the pipeline stages, we minimize the overall latency of the data processing, ensuring that each primitive is processed as soon as it arrives, without waiting for the entire event to be collected.

By applying both the \texttt{\#pragma HLS PIPELINE} and \texttt{\#pragma HLS DATAFLOW} directives, we ensure that each operation could process new data every clock cycle, allowing continuous data flow without waiting for the entire event to be read, thus reducing latency.

\begin{figure}[h!]
\begin{center}
\includegraphics[width=\linewidth]{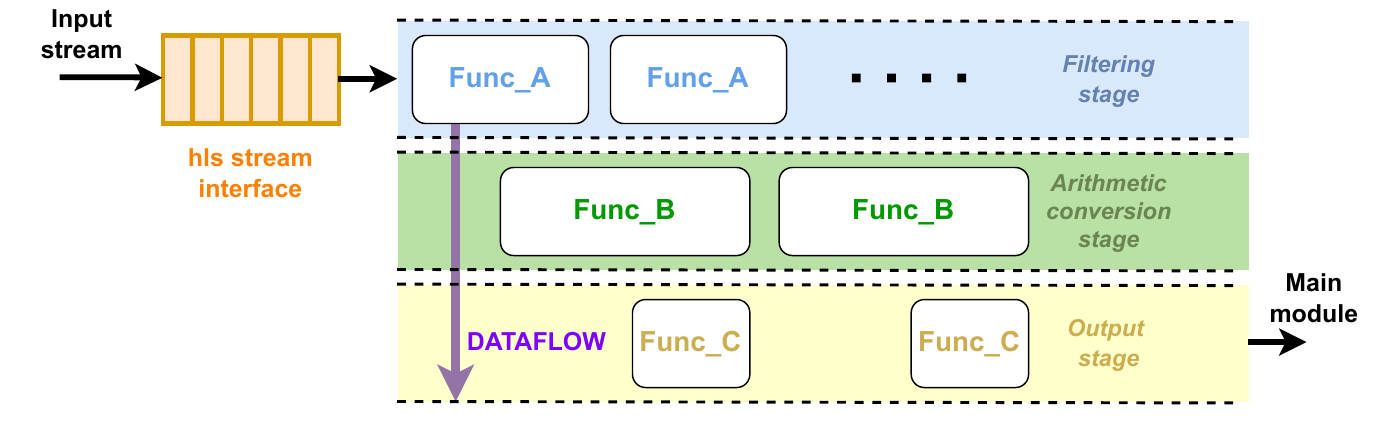}
\end{center}
\caption{Pipeline design of the DT module converter, utilizing a streamed interface inferred by the \texttt{hls::stream} class. The functions are executed in parallel following the HLS DATAFLOW paradigm, ensuring efficient data processing and throughput.}\label{fig:dtprocessor_stream}
\end{figure}

The pipelining technique has been applied in the DT module to demonstrate its effectiveness. By processing each primitive individually as it arrived, and using the \texttt{\#pragma HLS DATAFLOW} directive to allow concurrent execution of pipeline stages, we avoid the latency and resource demands associated with traditional deserialization and collective processing of data frames. The pipeline stages described above enabled continuous processing, achieving high throughput with minimal delay, as shown in Figure~\ref{fig:dtprocessor_stream}. We achieve significant reductions in latency (factor 2.43, see Section~\ref{sec:results}) while ensuring that data could be processed continuously and efficiently within the CMS Level-1 trigger system's stringent timing constraints.

%By leveraging pipelining and the \texttt{\#pragma HLS DATAFLOW} directive, we achieved significant reductions in latency while ensuring that data could be processed continuously and efficiently within the CMS Level-1 trigger system's stringent timing constraints.

\subsubsection{Pause points}

Throughout the development process, we implement pause points to verify the correctness and performance of the algorithm at various stages. These pause points are strategically placed after each implementation step to ensure that any issues could be identified and addressed promptly, preventing the propagation of errors to subsequent stages. 

This structured approach to validation made us realize the potential for creating a pipeline where automatic verification could be applied at each step. The concept of continuously verifying the design at each stage naturally evolved into the idea of integrating automation throughout the entire development process.

\section{Rapid prototyping workflow through automation scripting} \label{sec:automation}

\begin{figure}[h!]
\begin{center}
\includegraphics[width=\linewidth]{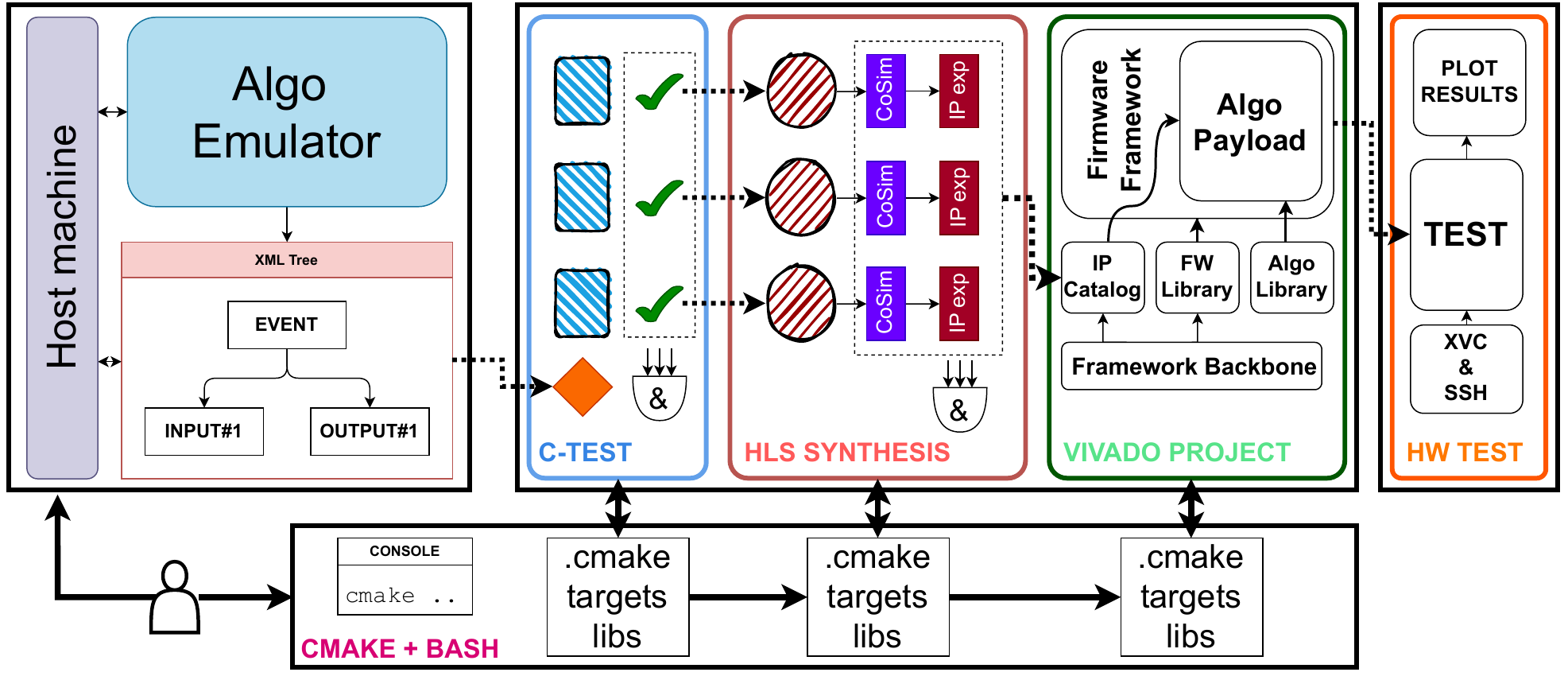}    
\end{center}
\caption{HLS build pipeline overview. The process begins with test vector generation in the algorithm emulator software. Next, the HLS modules are tested and synthesized. Following this, the board framework is built, and the top file along with the block design, including all HLS modules, is generated and implemented. Finally, after generating the bitstream, the design is tested on the target hardware.}\label{fig:build_pipeline}
\end{figure}

To manage the complexity of the development process and ensure efficient verification, we implement a custom local building and verification pipeline, see Figure~\ref{fig:build_pipeline}. This pipeline automates the compilation, testing, synthesis, and simulation steps, integrating them into a cohesive workflow that streamlines the development of the hardware-accelerated algorithm.

The validation of each step of the pipeline is crucial to ensure that the hardware-accelerated implementation of the track-finding algorithm functions correctly and mirrors the expected behavior of its software counterpart. The validation process involves several key steps:

\begin{enumerate}
    \item \textbf{Functional equivalence testing:} We perform comprehensive tests to compare the outputs of the hardware-implemented algorithm with those of the software version using identical input data from the CMS experiment. This ensures that the logic and computational integrity of the algorithm are preserved during the translation from software to hardware.

    \item \textbf{Bit-true verification:} Utilizing co-simulation features provided by HLS tools, we verify that the hardware design operates correctly at the bit level. This involves checking that the numerical computations and data handling in the FPGA match the software algorithm's precision and accuracy.

    \item \textbf{Iterative refinement and debugging:} Any discrepancies or unexpected behaviors observed during the validation steps of the pipeline are addressed through iterative refinement. This process involves debugging the hardware design, adjusting optimization directives in HLS, and re-validating until the hardware implementation consistently produces the correct results.
\end{enumerate}

Successful validation is achieved when the hardware implementation:

\begin{itemize}
    \item Produces output data that matches the software algorithm's results within error margins below 5\%.
    \item Operates correctly at the required high frequencies without data loss or corruption.
    \item Meets the performance targets for processing speed and latency improvements.
    \item Integrates seamlessly into the existing data processing infrastructure, facilitated by the automated prototyping workflow.
\end{itemize}

\subsection{C Simulation and unit testing}

After designing all the modules, we initiate the C simulation phase to evaluate their behavior. To automate this process, we implement a cmake target system that first compiles all the necessary libraries and source files. Cmake scripts are used to define build targets, manage dependencies, and configure the build environment.

We develop a unit testing system to verify each module individually. The testing process involves the following steps:

\begin{enumerate}
    \item \textbf{Extraction of test events}: We extract XML files from the emulator's algorithm simulation containing the expected test events, inputs, and outputs for each module across a number of events.
    \item \textbf{Creation of test targets}: For each module, a test target is created within the cmake system. These targets utilize a test template that is customized for each module by substituting placeholder names with the actual module names.
    \item \textbf{General Testing Library}: A general testing library developed in C++ is used to facilitate testing. This library consists of two main components:
    \begin{itemize}
        \item \textbf{CLIUtility module}: Responsible for processing input arguments of each test, reading the XML events, and managing command-line interfaces.
        \item \textbf{IModule library}: Contains a \texttt{Module\_IModule} file for each module, which receives the XML events and converts them into the expected input formats required by the module. It includes testing functions such as input conversion, output conversion (from C++ data types to arbitrary-precision types used in HLS), module execution, result comparison, and logging. Additionally, the \texttt{IModule} library includes a threshold-based verification system. A parameterized threshold is provided for each module to account for acceptable discrepancies in the output, arising from fixed-point arithmetic and hardware constraints. The results are verified by comparing them against the expected outputs within this threshold.
    \end{itemize}
    \item \textbf{Running the tests}: The tests are executed, with each module being fed by the corresponding inputs and producing outputs that are compared against the expected results from the XML files.
    \item \textbf{Recording results}: Test results are stored in a JSON file that records the events passed, events failed, and their respective inputs and outputs. This detailed record facilitates debugging and further analysis.
\end{enumerate}

\subsection{HLS synthesis and co-simulation}
\begin{figure}[h!]
\begin{center}
\includegraphics[width=\linewidth]{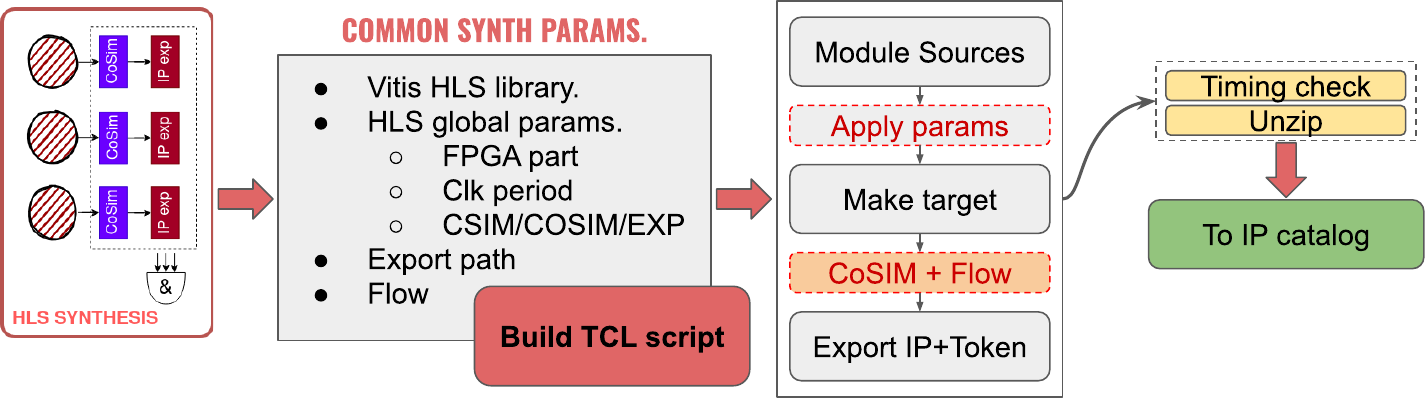}
\end{center}
\caption{Overview of the HLS synthesis and export process. The pipeline begins by setting common synthesis parameters such as FPGA part, clock period, and simulation/export options. A tcl script is generated to automate the process, which includes compiling the module sources, applying parameters, and targeting the design for synthesis. Co-simulation is performed along with the flow execution, and the IP, along with a token, is exported. The token ensures the IP is not rebuilt if the process is repeated. After a timing check and extraction, the IP is added to the catalog.}\label{fig:hls_pip_breakdown}
\end{figure}

Once the unit tests are passed, the synthesis of the HLS modules is performed. Figure \ref{fig:hls_pip_breakdown} shows the detailed parts of the HLS synthesis step. The pipeline includes: 1) \textbf{Synthesis}: Using the cmake system and tcl scripts, the HLS synthesis process is automated for each module. If any error is detected during synthesis, the pipeline stops, and the developer is prompted to refactor the code to resolve the issues. 2) \textbf{Co-simulation}: For each module, co-simulation is performed using the passed events from the JSON files as inputs. A co-simulation template is used to generate the necessary test benches and scripts. The co-simulation compares the C++ simulation results with the RTL simulation results to verify that the synthesized hardware matches the expected behavior.

%\begin{enumerate}
%\item \textbf{Synthesis}: Using the cmake system and tcl scripts, the HLS synthesis process is automated for each module. If any errors are detected during synthesis, the pipeline stops, and the developer is prompted to refactor the code to resolve the issues.
%\item \textbf{Co-simulation}: For each module, co-simulation is performed using the passed events from the JSON files as inputs. A co-simulation template is used to generate the necessary test benches and scripts. The co-simulation compares the C++ simulation results with the RTL simulation results to verify that the synthesized hardware matches the expected behavior.
%\end{enumerate}

\subsection{Integration}

Following the successful synthesis and verification of the HLS modules, the next critical step in the development pipeline is the integration of the algorithm into the hardware framework and the generation of the FPGA bitstream. This process involves combining our custom HLS design with existing hardware frameworks used at CERN, configuring communication protocols, and automating the build process through scripting.

At CERN, several hardware frameworks are employed to facilitate communication between FPGA boards and manage essential system functions. These frameworks provide standardized interfaces, communication protocols, and infrastructure components necessary for the operation of the boards. Key features of the hardware framework include:

\begin{figure}[h!]
\begin{center}
\includegraphics[width=\linewidth]{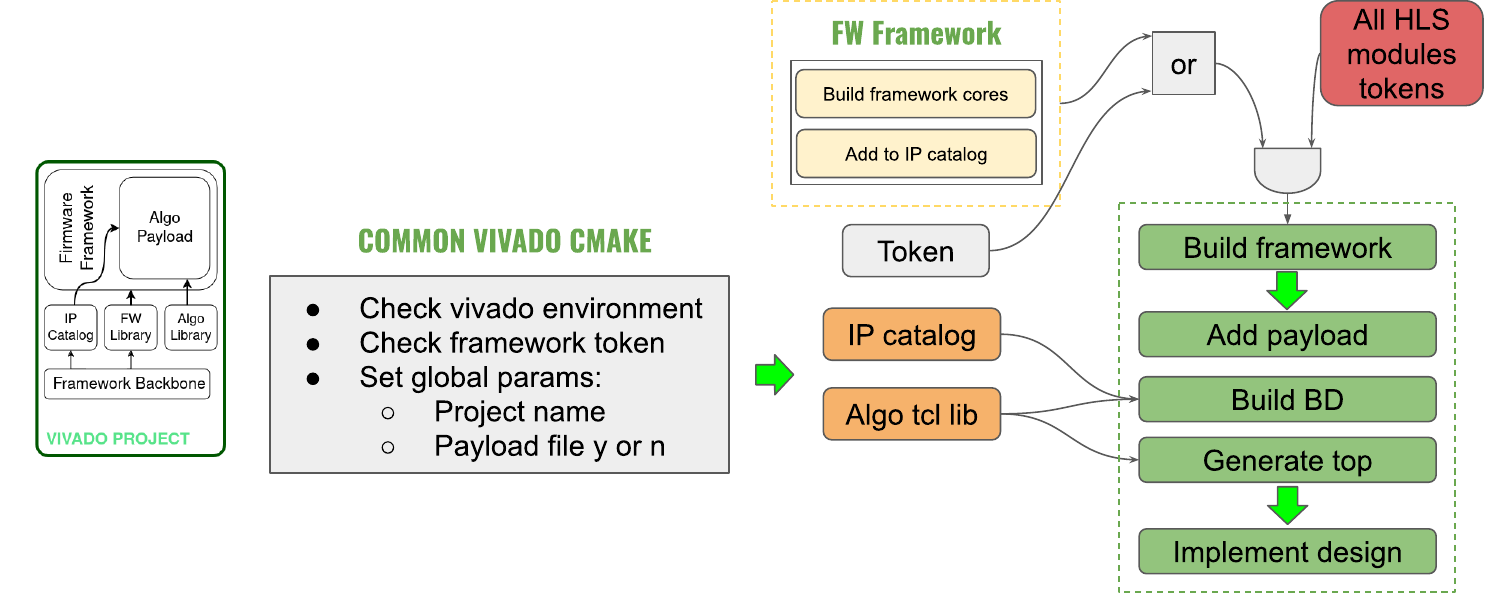}
\end{center}
\caption{Vivado project build flow. The process starts with either framework cores or HLS module tokens. The framework is then built, followed by adding the payload and constructing the block design (BD). The top-level design is generated, and the complete design is implemented. Tokens are used to avoid rebuilding modules unnecessarily. IPs are pulled from the IP catalog and supported by the algorithm tcl library, streamlining integration into the project.}\label{fig:vivado_pip}
\end{figure}

\begin{itemize}
    \item \textbf{Communication protocols}: Defines the methods for data exchange between boards, including the number of high-speed transceivers (e.g., Multi-Gigabit Transceivers or MGTs) used for high-throughput data links.
    \item \textbf{Control registers}: Sets up registers accessible from the processor system (e.g., embedded ARM cores), allowing for configuration, monitoring, and control of the FPGA logic.
    \item \textbf{Infrastructure components}: Includes clock management, reset logic, and other essential services required for stable operation.
\end{itemize}

We begin by building the hardware framework, ensuring that it is correctly configured to meet the requirements of our application. This involves selecting the appropriate number of transceivers, setting up control interfaces, and integrating any necessary communication protocols.

\subsection{Integration of HLS design}

Once the hardware framework is established, we proceed to integrate our synthesized HLS modules into the design. Figure \ref{fig:vivado_pip} shows the project building process. This process is automated using tcl scripting within Vivado, enabling us to build the entire block design and manage connections efficiently.

\begin{enumerate}
    \item \textbf{Importing HLS IP cores}: The HLS synthesis process generates IP cores for each module, which are added to the Vivado IP repository. These IP cores encapsulate the functionality of our algorithm modules and are ready for integration into the block design.
    \item \textbf{Automated block design creation}: Tcl scripts are used to instantiate the HLS IP cores and other necessary blocks within the Vivado block design environment. The scripts handle the placement of modules, configure parameters, and establish connections between blocks.
    \item \textbf{Connecting modules}: The scripts read the expected ports of both the hardware framework and the HLS payload modules, wiring them accordingly. This includes connecting data paths, control signals, and clock domains as required.
\end{enumerate}

After assembling the block design, a top-level HDL file is generated to serve as the entry point for synthesis and implementation. The top-level file defines the interface of the FPGA design, including all input and output ports, and ensures that all modules are correctly interconnected.

\begin{itemize}
    \item \textbf{Tcl scripting for top file generation}: We utilize tcl scripts to automate the generation of the top-level file. The scripts parse the port definitions from the hardware framework and the HLS modules, ensuring that all connections are accurately represented.
    \item \textbf{Interface definitions}: The top-level file specifies the physical interfaces, such as high-speed transceivers, GPIOs, and clock inputs, aligning them with the FPGA's pins and the board's connectors.
\end{itemize}

With the block design and top-level file prepared, we proceed to the synthesis and implementation stages:

\begin{enumerate}
    \item \textbf{Synthesis}: The combined design is synthesized to convert the RTL code into a gate-level netlist. This process involves optimizing the logic, mapping the design to the FPGA's resources, and ensuring that timing constraints are met.
    \item \textbf{Implementation}: The synthesized netlist undergoes placement and routing to assign the logic elements to specific locations on the FPGA and establish physical connections. This step is critical for meeting timing requirements and optimizing performance.
\end{enumerate}

To achieve optimal performance and ensure the design meets the operational requirements, we apply specific constraints during the implementation:

\begin{itemize}
    \item \textbf{Physical constraints}: Tcl scripts are used to generate a constraints file (\texttt{.xdc}) that specifies the placement of critical logic elements. This is particularly important when the FPGA is heavily utilized, or when certain paths require minimized propagation delays.
    \item \textbf{Timing constraints}: We define timing constraints to guide the synthesis and implementation tools in meeting the desired clock frequencies and setup/hold times.
    \item \textbf{Resource utilization constraints}: Limits are set on the usage of FPGA resources such as LUTs, BRAMs, and DSP slices to prevent over-utilization and ensure reliable operation.
\end{itemize}

The final step in the pipeline is the generation of the FPGA bitstream:

\begin{enumerate}
    \item \textbf{Bitstream file creation}: The implementation process produces a bitstream file (\texttt{.bit}) that encapsulates the configured FPGA design.
    \item \textbf{Verification}: Before deploying the bitstream, we perform a final verification step to ensure that all constraints are met and that there are no critical warnings or errors.
    \item \textbf{Deployment}: The bitstream is then ready to be loaded onto the FPGA for hardware testing and deployment within the CMS Level-1 trigger system.
\end{enumerate}

With the bitstream fully generated, the development pipeline concludes, and the design is ready for deployment and hardware testing. This comprehensive and automated pipeline—from algorithm preparation to bitstream generation—ensures that the hardware-accelerated algorithm is efficiently developed, thoroughly tested, and optimally implemented for operation within the CMS Level-1 trigger system. The automation of the hardware integration process and the application of the constrains, maximizes the performance and reliability of the FPGA implementation, meeting the rigorous demands of HEP data processing at the High-Luminosity Large Hadron Collider.

\subsection{From Local Automation to Future CI/CD Integration}

The entire hardware integration and bitstream generation process is automated within our build pipeline using tcl scripting and the capabilities of Vivado. This automation ensures consistency, reduces the potential for human error, and accelerates the development cycle.

It is important to note that the pipeline described above is developed for local development purposes. This setup allows individual developers or small teams to efficiently test and iterate on the design within their own development environments. The local pipeline provides the flexibility to make rapid changes, test new ideas, and perform detailed debugging without the overhead of a larger, more complex system.

To support repository management, scalability, and collaboration across larger teams, integrating this local pipeline into a Continuous Integration/Continuous Deployment (CI/CD) system is essential. CI/CD systems automate the building, testing, and deployment processes, ensuring that code changes are consistently integrated and validated.

\begin{itemize}
    \item \textbf{Consistency}: CI/CD pipelines enforce consistent build and test procedures across all developers, reducing integration issues and ensuring that the codebase remains stable.
    \item \textbf{Scalability}: As the team grows, a CI/CD system can handle multiple developers working concurrently, managing merges, and detecting conflicts early.
    \item \textbf{Automated Testing}: Automated tests can be run on every commit or pull request, ensuring that new changes do not introduce regressions.
    \item \textbf{Traceability and documentation}: CI/CD systems provide logs and reports for each build and test cycle, aiding in traceability and compliance with development standards.
    \item \textbf{Deployment automation}: Streamlines the process of deploying updates to hardware in the field, reducing manual intervention and potential errors.
\end{itemize}

By integrating the local development pipeline into a CI/CD system, we can leverage tools such as GitLab CI/CD, Jenkins, or Travis CI to automate the entire workflow. This integration ensures that the development process is scalable, maintainable, and aligned with best practices for software and hardware development.

\section{Experimental results} \label{sec:results}

This document presents a comparative analysis of the HLS-based implementation of the Overlap Muon Track Finder (OMTF) algorithm on an older \textit{Virtex-7 XC7VX690T} FPGA versus a new implementation on the \textit{Virtex UltraScale+ XCVU13P} FPGA. The new implementation incorporates significant architectural improvements and includes additional logic due to the requirements of the Phase-II upgrade, which represent an increase in the volume and complexity of data processed. Moreover, the new implementation includes extrapolation logic leveraging DSPs for multiplication operations \cite{Leguina_2023}. The comparison is summarized in Table~\ref{tab:comparison}.

\begin{table}[ht]
    \centering
    \renewcommand{\arraystretch}{1.2}
    \setlength{\tabcolsep}{4pt}
    \begin{tabular}{>{\centering\arraybackslash\bfseries}m{3cm} c|c}
        \hline\hline % Double line for the top
        \textbf{Parameter} & \textbf{\cite{zatbolony_hls}} & \textbf{Our Work} \\
        \hline % Single line for the bottom of the header row
        \textbf{Device} & XC7VX690T & XCVU13P \\
        \hline
        
        Frequency (MHz) & 160  & 360  \\
        Latency (cycles) & 54 & 50 \\
        Latency (ns) & 337.5  & 139  \\
        \hline
        
        LUT  & 123,964 (28.6\%) & 204,300 (11.8\%) \\
        FF  & 112,240 (12.9\%) & 198,022 (5.7\%) \\
        BRAM  & 720 (24.5\%) & 274 (10.2\%) \\
        DSP  & - & 204 (1.7\%) \\
        \hline\hline % Double line for the bottom
    \end{tabular}
    \caption{Comparison of key metrics for HLS implementations.}
    \label{tab:comparison}
\end{table}

The reductions in latency and resource usage, alongside the ability to handle more data, mark significant advancements in the algorithm's design.  Despite handling more data, the new implementation achieves a real-time latency improvement by a factor of:

\[
\text{Gain Factor} = \frac{337.5 \text{ ns (Virtex-7)}}{139 \text{ ns (Virtex UltraScale+)}} \approx 2.43
\]

This demonstrates the substantial advantage of the newer FPGA architecture combined with enhanced algorithm design. 

%Firmware / emulator comparisons are also measured and the agreement is higher than 90\% mostly due to bit-error mismatches. % TODO!!! 

\section{Discussion} \label{sec:discussion}

%\subsection{Interpretation of results}

The results are expected to demonstrate that hardware acceleration using High-Level Synthesis (HLS) significantly improves processing speeds by 2.25 and reduces latency by factor 2.43 in a high-frequency application such as the muon track-finding algorithm of the CMS Level-1 trigger system. The successful implementation confirms that HLS is an effective tool for translating complex algorithms into efficient hardware designs suitable for real-time applications in HEP. The performance gains validate the optimization techniques applied, such as parallel processing and pipelining, highlighting their impact on enhancing computational efficiency without sacrificing accuracy.

%\subsection{Implications for HEP research}

The advancements achieved through this work have substantial implications for HEP research. By enabling real-time data processing with improved accuracy, researchers can explore new physics phenomena that require rapid and precise measurements, such as the detection of rare particles or events occurring at high luminosities. The adoption of hardware acceleration expands the capabilities of trigger systems, potentially leading to more effective data collection strategies and enhancing the overall scientific output of experiments like CMS.

%\subsection{Scalability and future directions}

The methodologies developed in this study demonstrate scalability to other algorithms and detector systems within HEP. Future work may involve extending the use of HLS and hardware acceleration to additional components of the trigger system or other experiments facing similar computational challenges. Exploring hardware/software co-design approaches and integrating machine learning algorithms into hardware accelerators are promising directions. Continued development of HLS tools and techniques will further simplify the design process, making hardware acceleration more accessible to a broader range of researchers.

%\subsection{Conclusion}

This study highlights the transformative potential of hardware acceleration in HEP, showcasing how high-level synthesis can effectively bridge the gap between complex software algorithms and hardware implementation. By harnessing HLS, we achieve significant improvements in processing speed and latency reduction for the muon track-finding algorithm in the CMS experiment. The successful application of these techniques not only enhances current data analysis capabilities but also sets the stage for future innovations, enabling physicists to tackle increasingly complex challenges and drive groundbreaking discoveries in the field.

%There are 5 heading levels

%\subsection{Level 2}
%\subsubsection{Level 3}
%\paragraph{Level 4}
%b                                                                                                 n\subparagraph{Level 5}

%%Figures, tables, and images will be published under a Creative Commons CC-BY licence and permission must be obtained for use of copyrighted material from other sources (including re-published/adapted/modified/partial figures and images from the internet). It is the responsibility of the authors to acquire the licenses, to follow any citation instructions requested by third-party rights holders, and cover any supplementary charges.

\section{Additional Requirements}

For additional requirements for specific article types and further information please refer to \href{http://www.frontiersin.org/about/AuthorGuidelines#AdditionalRequirements}{Author Guidelines}.

\section*{Conflict of Interest Statement}
%All financial, commercial or other relationships that might be perceived by the academic community as representing a potential conflict of interest must be disclosed. If no such relationship exists, authors will be asked to confirm the following statement: 

The authors declare that the research was conducted in the absence of any commercial or financial relationships that could be construed as a potential conflict of interest.

\section*{Author Contributions}

The Author Contributions section is mandatory for all articles, including articles by sole authors. If an appropriate statement is not provided on submission, a standard one will be inserted during the production process. The Author Contributions statement must describe the contributions of individual authors referred to by their initials and, in doing so, all authors agree to be accountable for the content of the work. Please see  \href{https://www.frontiersin.org/about/policies-and-publication-ethics#AuthorshipAuthorResponsibilities}{here} for full authorship criteria.

\section*{Funding}
This work and the authors are partially supported by ERC grant (INTREPID, 101115353) and the Ministerio de Ciencia e Innovaci\'on (Spain) with the project PID2020-113341RB-I00. Funded by the European Union. Views and opinions expressed are however those of the author(s) only and do not necessarily reflect those of the European Union or the European Research Council Executive Agency. Neither the European Union nor the granting authority can be held responsible for them. 

\section*{Acknowledgments}
We would like to thank the Level-1 Trigger Project of the CMS experiment for they support to the authors. Special thanks to K. Bunkowski, P. Fokow, M Konecki, and W. Zabolotny for their insight in the HLS implementation of the OMTF algorithm, as well as to M. Batchis and his team for facilitating the validation of the algorithm.  

% Please see the availability of data guidelines for more information, at https://www.frontiersin.org/about/author-guidelines#AvailabilityofData

\bibliographystyle{Frontiers-Harvard} %  Many Frontiers journals use the Harvard referencing system (Author-date), to find the style and resources for the journal you are submitting to: https://zendesk.frontiersin.org/hc/en-us/articles/360017860337-Frontiers-Reference-Styles-by-Journal. For Humanities and Social Sciences articles please include page numbers in the in-text citations 
\bibliography{bibliography}

%%% Make sure to upload the bib file along with the tex file and PDF
%%% Please see the test.bib file for some examples of references

%%% Please be aware that for original research articles we only permit a combined number of 15 figures and tables, one figure with multiple subfigures will count as only one figure.
%%% Use this if adding the figures directly in the mansucript, if so, please remember to also upload the files when submitting your article
%%% There is no need for adding the file termination, as long as you indicate where the file is saved. In the examples below the files (logo1.eps and logos.eps) are in the Frontiers LaTeX folder
%%% If using *.tif files convert them to .jpg or .png
%%%  NB logo1.eps is required in the path in order to correctly compile front page header %%%

%%% If you don't add the figures in the LaTeX files, please upload them when submitting the article.
%%% Frontiers will add the figures at the end of the provisional pdf automatically
%%% The use of LaTeX coding to draw Diagrams/Figures/Structures should be avoided. They should be external callouts including graphics.

\end{document}